\documentclass[twocolumn,prl,showpacs,preprintnumbers,amsmath,amssymb]{revtex4}
\usepackage{graphicx}
\usepackage{dcolumn}
\usepackage{bm}

\begin{document}


\title{Doping Evolution of the Underlying Fermi Surface in La$_{2-x}$Sr$_x$CuO$_4$}

\author{T. Yoshida$^1$, X. J. Zhou$^2$, K. Tanaka$^1$, W. L. Yang$^2$, Z. Hussain$^3$, Z.-X. Shen$^2$, A.
Fujimori$^1$, Seiki Komiya$^4$, Yoichi Ando$^4$, H. Eisaki$^5$, T.
Kakeshita$^6$, S. Uchida$^7$} \affiliation{$^1$Department of
Complexity Science and Engineering, University of Tokyo, Kashiwa,
Chiba 277-8561, Japan} \affiliation{$^2$Department of Applied
Physics and Stanford Synchrotron Radiation Laboratory, Stanford
University, Stanford, CA94305} \affiliation{$^3$Advanced Light
Source, Lawrence Berkeley National Lab, Berkeley, CA 94720}
\affiliation{$^4$Central Research Institute of Electric Power
Industry, Komae, Tokyo 201-8511, Japan}
 \affiliation{$^5$National
Institute of Advanced Industrial Science and Technology, Tsukuba
305-8568, Japan} \affiliation{$^6$Superconductivity Research
Laboratory, ISTEC, Shinonome 1-10-13, Koto-ku, Tokyo 135-0062,
Japan} \affiliation{$^7$Department of Physics, University of
Tokyo, Bunkyo-ku, Tokyo 113-0033, Japan}
\date{\today}

\begin{abstract}
We have performed a systematic doping dependent study of
La$_{2-x}$Sr$_x$CuO$_4$ (LSCO) (0.03$\leq x \leq$0.3) by
angle-resolved photoemission spectroscopy. In the entire doping
range, the underlying ``Fermi surface" determined from the low
energy spectral weight approximately satisfies Luttinger's
theorem, even down to the lightly-doped region. This is in strong
contrast to the result on Ca$_{2-x}$Na$_x$CuO$_2$Cl$_2$ (Na-CCOC),
which shows a strong deviation from Luttinger's theorem. The
differences between LSCO and Na-CCOC are correlated with the
different behaviors of the chemical potential shift and spectral
weight transfer induced by hole doping.
\end{abstract}

\maketitle
The pseudo-gap behavior in the underdoped high-$T_c$ cuprates has
attracted significant attention and is one of the most challenging
problems in strongly correlated systems. As a result of pseudo-gap
opening around $k$=($\pi$,0), which is observed by angle-resolved
photoemission spectroscopy (ARPES) studies of underdoped cuprates
\cite{andrea}, the electronic specific heat $\gamma$
\cite{loram,momono,nakano} and the Pauli paramagnetic
susceptibility $\chi_s$ \cite{nakano} decrease with decreasing
hole concentration $x$. Such unconventional behaviors in the
underdoped cuprates are thought to be a remarkable example of
strong deviation from the normal Fermi liquid. However, it is
still unclear whether they can be understood starting from a Fermi
liquid or should be understood from a fundamentally new kind of
ground state.

Theoretically, two pictures in the underdoped cuprates, namely, a
large Fermi surface centered at ($\pi,\pi$) or a small Fermi
surface centered at ($\pi/2,\pi/2$) have been considered. In
theories starting from the Fermi liquid influenced by strong
antiferromagnetic fluctuations, the large Fermi surface has been
obtained \cite{Prelovsek}. Uniform resonant-valence-bond (RVB)
states \cite{fukuyama}, cellular dynamical mean field theory
(CDMFT) \cite{kotliar}, etc., also predict the large Fermi
surface. On the other hand, flux RVB and other kinds of exotic
symmetry breaking \cite{chakraverty,varma} lead to the small Fermi
surface. In the Fermi-liquid theory, the carrier number is given
by the Fermi surface volume (Luttinger's theorem). However, even
for the large Fermi surface, most of numerical calculations on the
Hubbard and $t$-$J$ models have predicted noticeable deviation
from Luttinger's theorem in the underdoped region that the
occupied area of the Fermi surface becomes significantly larger
than that expected from the hole concentration
\cite{Esirgen,Maier,Putikka}. Since the Fermi surface and its
Luttinger's theorem are at the center of the Fermi-liquid concept,
systematic experimental studies of the Fermi surface and its
volume change as a function of hole doping should be crucial for
understanding the ground state of the pseudo gap state in the
cuprates.

Recent ARPES studies on lightly-doped cuprates
\cite{Metallic,KMShen} have shown that the Fermi surface is
basically large and that, while the ($\pi$,0) region remains
(pseudo) gapped, a quasi-particle (QP) band is formed and crosses
the Fermi level ($E_F$) in the nodal (0,0)-($\pi$,$\pi$)
direction, leading to the picture that only part of the Fermi
surface survives as an ``arc" around the node. The QP crossing in
lightly-doped LSCO and moderately lightly-doped Na-CCOC is
certainly responsible for their metallic behavior \cite{ando}. The
volume of the ``Fermi surface" (which is defined under this
Fermi-arc/pseudo-gap situation) in Na-CCOC was found to strongly
deviate from Luttinger's theorem as the doping approaches zero
\cite{KMShen}, consistent with the theoretical predictions
\cite{Esirgen,Maier,Putikka}.

In the present study, the doping evolution of the unconventional
electronic structure of LSCO with the Fermi arc and the pseudogap
have been systematically studied by ARPES in the entire doping
range. We have revealed the evolution of the ``Fermi surface" with
hole doping and found that the Fermi surface area almost fulfills
Luttinger's theorem in the entire doping range even down to $x\sim
0.03$, namely, down to the ``spin-glass" region. This is in strong
contrast with Na-CCOC \cite{KMShen}. We shall discuss
phenomenological as well as microscopic origins of the differences
between the two systems.

\begin{figure}
\includegraphics[width=8.5cm]{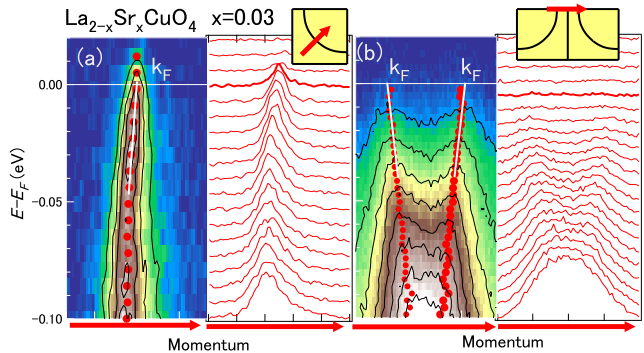}
\caption{\label{NodeAntiNode}(Color online) ARPES spectra of
La$_{1.97}$Sr$_{0.03}$CuO$_4$. (a) Intensity plot in $E$-$k$ space
(left) and MDC's (right) for the nodal cut (inset). (b) Intensity
plot in $E$-$k$ space (left) and MDC's (right) for the anti-nodal
cut near ($\pi$,0) (inset). Red dots represent the peak positions
in MDC's. Even in the pseudogap region near ($\pi$,0), one can
determine $k_F$ by extrapolating the MDC peaks to $E_F$.}
\end{figure}

\begin{figure*}
\includegraphics[width=18cm]{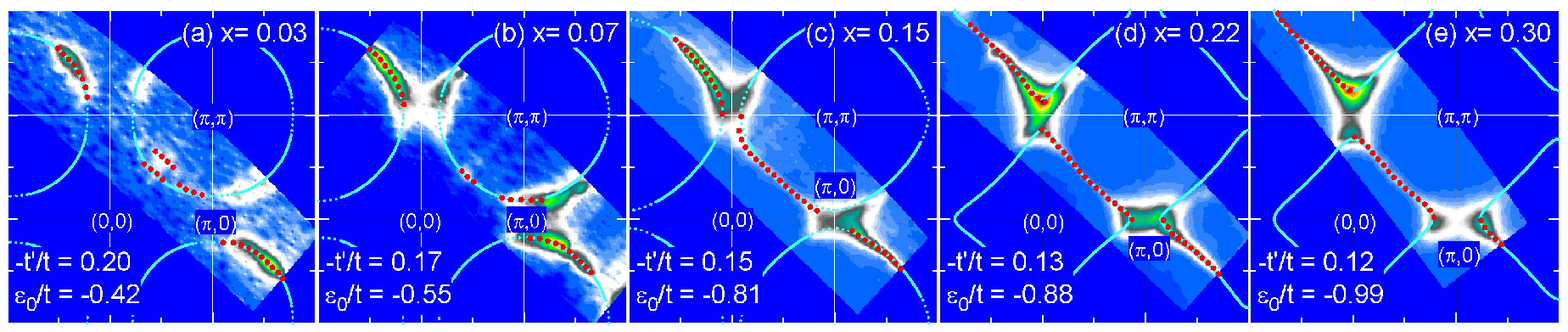}
\caption{\label{FS}(Color) Spectral weight mapping in $k$-space at
$E_F$ in La$_{2-x}$Sr$_x$CuO$_4$. Red dots indicate $k_F$
positions determined by the MDC peaks at $E_F$ (see
Fig.\ref{NodeAntiNode}). The blue curves show the Fermi surface
interpolated by the tight-binding model.}
\end{figure*}

The ARPES measurements were carried out at BL10.0.1 of Advanced
Light Source, using incident photons of 55.5 eV. We used a SCIENTA
SES-2002 analyzer with the total energy resolution of 20 meV and
the momentum resolutions of 0.02$\pi$ in units of 1/$a$, where
$a$=3.8 \textrm{\AA} is the lattice constant. High-quality single
crystals of LSCO were grown by the traveling-solvent floating-zone
method. The critical temperature ($T_c$) of $x$ = 0.07, 0.15 and
0.22 samples were 14, 41 and 22 K, respectively, and $x$ = 0.03
and 0.30 samples were non-superconducting. The samples were
cleaved \textit{in situ} and measurements were performed at 20 K
as in the previous studies \cite{yoshidaOD}. In the present
measurements, the electric vector $\mathbf{E}$ of the incident
photons lies within the CuO$_2$ plane, 45 degrees rotated from the
Cu-O direction and is parallel to the Fermi surface segment around
the diagonal region. This measurement geometry enhances dipole
matrix elements in this $ \mathbf{k}$ region because the wave
function has $x^2-y^2$ symmetry \cite{yoshidaOD}.

\begin{figure*}
\includegraphics[width=18cm]{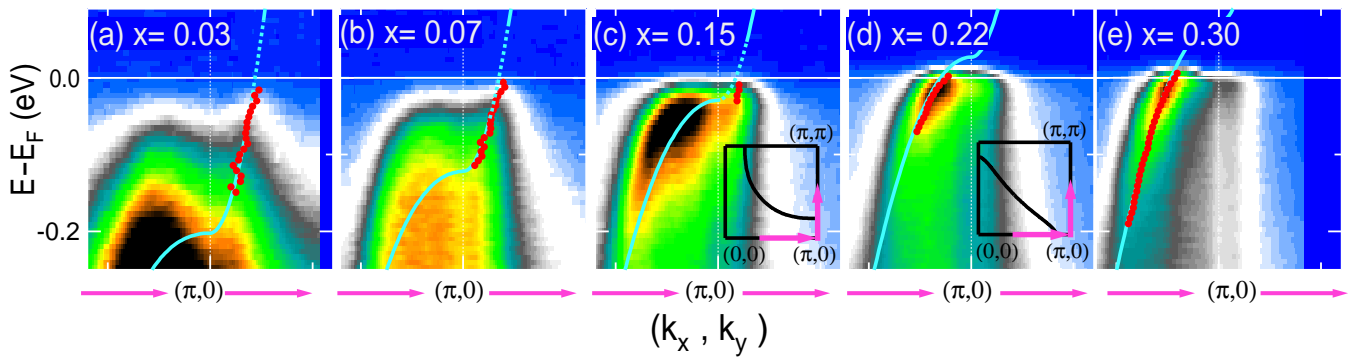}
\caption{\label{disp}(Color) Intensity plot in $E$-$k$ space along
symmetric lines (0,0)-($\pi$,0)-($\pi,\pi$) in
La$_{2-x}$Sr$_x$CuO$_4$. The direction and length of the arrows in
the inset correspond to the horizontal axis of the color plots.
The blue lines show the tight-binding interpolation. Red dots
indicate MDC peaks.  }
\end{figure*}

Figure \ref{NodeAntiNode} shows ARPES results for LSCO ($x$=0.03),
illustrating how the ``Fermi surface crossing" in the pseudo-gap
state can be determined from the momentum distribution curves
(MDC's). In the nodal direction [(a)], where there is no gap, the
underlying Fermi surface is easily determined by the MDC peak
position at $E_F$. Near the ($\pi$,0) point [(b)], on the other
hand, the spectral weight is strongly suppressed due to the
pseudo-gap formation. However, as shown in panel (b), one can
determine $k_F$ by extrapolating the MDC peaks up to $E_F$ even
when the spectral weight is suppressed toward $E_F$. Hereafter, we
refer to the Fermi surface thus determined in the pseudo-gap state
as the ``underlying" Fermi surface. This underlying Fermi surface
is consistent with the strong temperature dependence of Hall
coefficient $R_H$ \cite{hwang,AndoHall} if the low-energy spectral
weight around ($\pi$,0) does not contribute to charge transport at
low temperatures but does at high temperatures \cite{AndoHall}.

Accordingly, we have determined the (underlying) Fermi surfaces
for the entire doping range as shown in Fig. \ref{FS}. Here, red
dots in the momentum space indicate the $k_F$ position of the
(underlying) Fermi surface determined using MDC's. Spectral weight
at $E_F$ is also mapped in color plot. While the optimum and
overdoped samples ($x \geq$0.15) show strong intensities
throughout the entire Fermi surface, the underdoped samples ($x
\leq$0.1) show weak or suppressed spectral weight around
($\pi$,0), i.e., a ``truncated" Fermi surface or a Fermi ``arc"
due to the pseudo-gap formation around ($\pi$,0). With the $k_F$
points in the first and second Brillouin zone (BZ) as well as the
shadow band in the first BZ \cite{SB}, we could precisely
determine the absolute momentum position of the Fermi surface for
the entire doping range. From the $E$-$k$ space ARPES intensity
plots shown in Fig. \ref{disp}, one can clearly see the doping
dependence of the pseudogap opening around ($\pi$,0). Although the
spectral intensity at $E_F$ around ($\pi$,0) becomes weak for
$x\leq$ 0.15 because of the pseudo-gap or superconducting gap
opening, it is possible to identify the underlying band
dispersions from the MDC peak positions (red dots) as described
above. Thus, one can determine the $k_F$ around ($\pi$,0) even
near the crossover from the hole-like Fermi surface to the
electron-like Fermi surface.

The $k_{F}$'s determined in Figs. \ref{FS} have been fitted to the
single-band tight-binding (TB) model
$\varepsilon_k=\varepsilon_0-2t(\cos k_xa +\cos k_ya)
-4t^\prime\cos k_xa \cos k_ya \nonumber-2t^{\prime\prime}(\cos
2k_xa +\cos 2k_ya)$, as shown by blue curves. Here, $t$,
$t^\prime$ and $t^{\prime\prime}$ are the first, second and third
nearest neighbor transfer integrals between Cu sites. We have
assumed constant $t=0.25$ eV and relationship
$-t^{\prime\prime}/t^\prime=1/2$ for all the doping levels, and
regarded $t^\prime$ and $\varepsilon_0$ as adjustable parameters.
The fitting results of the TB parameters, $-t^\prime/t$ and
$\varepsilon_0$ are shown in each panel of Fig. \ref{FS}. Although
the absolute values of $t$, $t^\prime$ and $t^{\prime\prime}$ are
smaller than those determined by the band structure calculation by
a factor of $\sim$ 0.5, the relative magnitude of the TB
parameters agree rather well with the band-structure calculation,
e. g. $-t^\prime/t\sim$ 0.15 \cite{pavarini}. In the overdoped
region, the entire dispersion is almost perfectly fitted by the TB
model, although some misfit can be seen for the kink structure due
to phonons (particulary in the nodal kink) \cite{lanzara,
zhouNature} and the extremely flat band dispersion around
($\pi$,0) in the underdoped region. Here, the best-fit TB
parameters $-t^\prime/t$ shown in Fig. \ref{FS} exhibit a clear
doping dependence. The increase of $-t^\prime/t$ with decreasing
$x$ can be explained by the increase of the Cu-apical oxygen
distance with decreasing $x$ \cite{radaelli} according to
Ref.\cite{pavarini}. Also, it is consistent with the CDMFT
calculation \cite{kotliar} which indicates the increasing
correlation effects in the underdoped region.

\begin{figure}
\includegraphics[width=8.5cm]{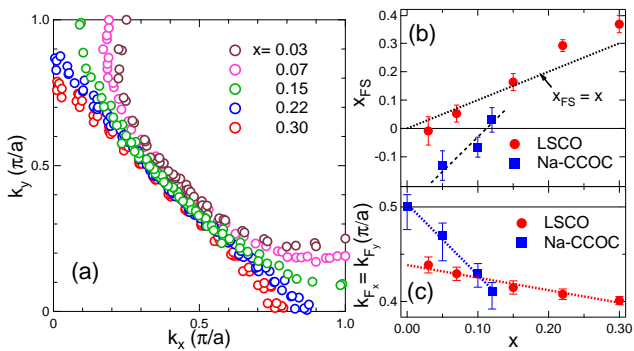}
\caption{\label{Luttinger}(Color online) Doping dependence of the
Fermi surface in LSCO. (a) $k_F$ position for each doping
determined by MDC peaks at $E_F$. (b) Doping dependence of the
hole number $x_{FS}$ deduced from the Fermi surface area.
Luttinger's theorem $x_{FS}=x$ is shown for comparison (broken
lines). (c) Doping dependence of the $k_F$ position in the node
direction. Data for Na-CCOC \cite{KMShen} are also plotted.}
\end{figure}

Figure \ref{Luttinger} (a) summarizes the experimental Fermi
surfaces. The hole number deduced from the area of the
experimental Fermi surface, $x_{FS}$, obtained assuming
Luttinger's theorem is plotted in Fig. \ref{Luttinger} (b) as a
function of $x$. It is remarkable that $x_{FS}$ approximately
accords with that predicted by Luttinger's theorem (broken lines)
in the entire doping range. This result is in strong contrast with
the result on Na-CCOC \cite{KMShen}, where the $x_{FS}$ shows a
strong deviation from Luttinger's theorem in the underdoped
region, as also shown in Fig. \ref{Luttinger}(b). [However, a
closer inspection reveals that there may be a slight deviation
from Luttinger's theorem in the lightly doped LSCO ($x$=0.03) in
the same direction as Na-CCOC.]

In order to highlight the different shapes of the Fermi surface
between LSCO and Na-CCOC, we have also plotted the doping
dependence of the $k_F$ position in the nodal direction in Fig.
\ref{Luttinger}(c). Extrapolation of $k_F$ to $x$=0 gives
($\pi$/2,$\pi$/2) for Na-CCOC, probably reflecting the situation
where the low energy excitation in the low-doping limit comes from
(the tail of) the top of the lower Hubbard band (LHB), which has
the band maximum at ($\pi$/2,$\pi$/2), on the boundary of the
antiferromagnetic BZ \cite{KMShenQP,KMShen}. In contrast, the
$k_F$ in LSCO for $x$=0.03 is still away from ($\pi$/2,$\pi$/2).
This would be related to the observation that the spectral weight
at $E_F$ comes from the QP peak which is well separated from the
LHB \cite{Metallic}. While the LHB in Na-CCOC approaches $E_F$
with hole doping \cite{KMShenQP}, that in LSCO stays away from
$E_F$ \cite{Metallic} and the spectral weight is transferred to
the QP band near $E_F$. This contrasting behavior between LSCO and
Na-CCOC is closely related to the different chemical potential
shifts, that is, while photoemission spectra of Na-CCOC show rigid
shifts with hole doping \cite{KMShenQP}, those of LSCO show slow
shifts reflecting the pinning of chemical potential at in-gap
states \cite{inoCP}. Although the large Fermi surface is observed
in Na-CCOC, the doping evolution of the electronic structure is
somewhat similar to the small Fermi surface picture in the sense
that the chemical potential appears to be shifted from the top of
the LHB downward. The large deviation from Luttinger's theorem may
be reminiscent of the small Fermi surface behavior.

Recently, the LHB of underdoped La$_2$CuO$_4$ and
Ca$_2$CuO$_2$Cl$_2$ have been interpreted as polaronic side bands
\cite{KMShenQP, Rosch}. Although this picture well explains the
LHB feature of the undoped samples, it is not straightforward to
understand the different doping evolution between LSCO and Na-CCOC
mentioned above. In the polaron picture, the peak of the LHB is
shifted toward the $E_F$ with hole doping since electron-phonon
coupling is weakened by screening effects. However, the peak of
the LHB in LSCO stays at almost the same binding energy with hole
doping. This suggests that in LSCO the local charge density does
not change with hole doping, reminiscent of a phase separation
between hole-poor and hole-rich region. Indeed, the chemical
potential pinning observed for LSCO \cite{inoCP} is a natural
consequence of a mixture of different hole concentrations.
According to a recent theoretical study, a mixed phase of
antiferromagnetic (AF) and superconducting (SC) states is proposed
to explain the coexistence of the QP states and the LHB in LSCO
\cite{Dagotto}, and captures the characteristic two-component
behavior of the ARPES results of underdoped LSCO
\cite{inoPRB,Metallic}. Also, it has been predicted that phase
separation between the insulating and metallic phases occurs under
a certain regime of electron-phonon coupling strength
\cite{capone}. Thus, observed spectral features for LSCO are
suggestive of phase separation picture. This inference is
collaborated by the transport evidence for a phase separation in
lightly-doped LSCO \cite{AndoMagnet}. For a phase-separated state,
one would normally expect to observe two Fermi surfaces
corresponding the two phases. If the difference of the hole
concentration between the AF and SC phases is not so large or if
there are temporal fluctuations between the two phases, however,
only the average of the Fermi surfaces of the two phases would be
observed and would approximately fulfill Luttiger's theorem.
Alternatively, the observed Fermi surface may come only from the
hole-rich region. If the occupied area of the underlying Fermi
surface becomes larger than that expected from the hole
concentration as theoretically predicted
\cite{Esirgen,Maier,Putikka}, this effect and the increase of the
hole concentration in the hole-rich region may cancel each other
and Luttinger's theorem may be accidentally fulfilled.

In summary, we have observed systematic changes of the underlying
Fermi surfaces in LSCO over a wide doping range. The area of the
obtained underlying Fermi surface approximately satisfies
Luttinger's theorem even in the lightly-doped region. This
behavior is contrasted with that of Na-CCOC, which shows a clear
deviation from Luttinger's theorem. Possible origins of the
difference between LSCO and Na-CCOC have been discussed in
relation to the chemical potential pinning and the phase
separation in the underdoped LSCO.

We are grateful to P. Prelovsek, N. Nagaosa, C. M. Ho and M. Ido
for enlightening discussions. This work was supported by a
Grant-in-Aid for Scientific Research in Priority Area ``Invention
of Anomalous Quantum Materials", Grant-in-Aid for Young Scientists
from the Ministry of Education, Science, Culture, Sports and
Technology. ALS is operated by the Department of Energy's Office
of Basic Energy Science, Division of Materials Science.


\end{document}